# Challenging Fuel Cycle Modeling Assumptions: Facility and Time Step Discretization Effects

Robert W. Carlsen (rwcarlsen@gmail.com), Paul P.H. Wilson
University of Wisconsin – Madison, Department of Nuclear Engineering and Engineering Physics, Madison, WI 53706

## Abstract

Due to the diversity of fuel cycle simulator modeling assumptions, direct comparison and benchmarking can be difficult. In 2012 the Organisation for Economic Co-operation and Development (OECD) completed a benchmark study that is perhaps the most complete published comparison performed. Despite this, various results from the simulators were often significantly different because of inconsistencies in modeling decisions involving reprocessing strategies, refueling behavior, reactor end-of-life handling, etc. This work identifies and quantifies the effects of selected modeling choices that may sometimes be taken for granted in the fuel cycle simulation domain. Four scenarios are compared using combinations of either fleet-based or individually modeled reactors with either monthly or quarterly (3-month) time steps. The scenarios approximate a transition from the current U.S. once-through light water reactor (LWR) fleet to a full sodium fast reactor (SFR) fuel cycle. The Cyclus fuel cycle simulator's plug-in facility capability along with its market-like dynamic material routing allow it to be used as a level playing field for comparing the scenarios. When under supply-constraint pressure, the four cases exhibit noticeably different behavior. Fleet-based modeling is more efficient in supply-constrained environments at the expense of losing insight on issues such as realistically suboptimal fuel distribution and challenges in reactor refueling cycle staggering. Finer-grained time steps also enable more efficient material use in supply-constrained environments resulting in much lower standing inventories of separated Pu. Large simulations with fleet-based reactors run much more quickly than their individual reactor counterparts. Gaining a better understanding of how these and other modeling choices affect fuel cycle dynamics will enable making more deliberate decisions with respect to trade-offs such as computational investment vs. realism.

# I INTRODUCTION

The diversity of assumptions embedded within various fuel cycle simulators pose challenges for direct comparison and benchmarking. In 2012 the Organisation for Economic Co-operation and Development (OECD) completed a benchmark study [1] that is perhaps one of the most complete published comparisons performed. Despite this, however, various results from the different simulators were often significantly different because of inconsistent modeling decisions involving reprocessing strategies, refueling behavior, reactor end-of-life handling, etc. This work attempts to identify and quantify the effects of selected modeling choices relative to facility modeling discretization (i.e., individual facilities vs. aggregate fleets) and time step duration. A better understanding of how these affect fuel cycle dynamics will allow for making more deliberate decisions with respect to trade-offs such as computational investment vs. realism.

This work identifies and quantifies the effects of selected modeling choices that may sometimes be taken for granted in the fuel cycle simulation domain. Four scenarios are compared using combinations of either fleet-based or individually modeled reactors with either monthly or quarterly (3-month) time steps. Time step duration and the degree facility discretization are design decisions that are handled in a variety of unique ways among nuclear fuel cycle simulators, and a better understanding of their effects can serve to increase confidence in results and can help make inter-simulator comparisons more meaningful. Reducing error and noise artifacts associated with modeling decisions will provide a more stable foundation for understanding the significance of external market, political, and social forces that often dominate the industry in the real world.

## I.A Motivation and Background

Informal comparison of results from the VISION and Cyclus simulators when running equivalent scenarios revealed some surprising discrepancies. One of the most apparent was a large difference in separated Pu levels between the two simulators — something that some in the DOE and nuclear community are very sensitive to. For one particular scenario, the Cyclus results also showed no fuel shortages for reactors, while the VISION results indicated noticeable shortages. This observation prompted a closer investigation to determine if the differences were caused by mistakes in simulator input, bugs in the simulator(s), or different (yet still reasonable) modeling choices of the simulators. This investigation eventually developed into this work which shows important effects of that are operative in many fuel cycle simulators.

Traditionally, many fuel cycle simulators have used a system dynamics approach [2] to modeling. These simulators model several *stocks* representing isotopes or isotope groups that have quantities that change over time because of *flows* between them. The flows are determined by equations that are a function of the state of the system (i.e., the system's stocks). A constructed system is then solved in discrete or continuous time depending on the limitations of the software and constraints imposed by the model. In system dynamics-based simulators, the number of stocks and potential flows between them is generally a static property of the modeled system, making it somewhat difficult to swap different reactor models in and out of the same scenario without wiring them into the simulator manually.

System dynamics-based fuel cycle simulators generally use a collection of stocks to represent the state of groups of like-facilities commonly referred to as "fleets". As the simulation steps through time, the levels of stocks (e.g., reactor fresh fuel inventory, repository waste inventories, etc.) are adjusted according to the calculated flow values. Fuel cycle simulators that model facility fleets built on a system dynamics foundation include VISION [3], DANESS [4], DYMOND, and ORION [5] among others. In fleet-based modeling:

- Groups of facilities (usually all facilities of a given type) are treated as a single, aggregate entity. Therefore, all facilities experience a single, identical history.
- Resource flows are modeled as continuous, within the limitations of time step discretization. Therefore, material is acquired and discharged in uniform amounts at every time step.

This work investigates in detail some differences between fleet vs. individual reactor modeling. The difference between these two modeling styles in terms of their effects on output of fuel cycle simulators in general is not well known. This analysis is designed to provide both quantitative and qualitative insight about how fleet-based continuous flow models differ from individual facility discrete flow models.

Many fuel cycle simulators step through time in discrete steps. The duration of a single time step can have significant impact on the dynamics of a simulation. The time step duration can affect the magnitude of individual material transfers between facilities and correspondingly affects the noise level in facility material inventories and flows. It can also act as a minimum bound for outage durations and is a minimum bound in general for state changes within the entire system. The effects of the time step duration are also investigated together with fleet vs. facility modeling.

Questions of interest include:

- What effects drive differences in fleet-based and individual-based facility modeling?
- How does changing the time step duration affect simulations?
- Does the impact of changing time step duration differ in fleet-based and individual-based facility modeling?

## I.B Cyclus

One of the major motivations for the development of the Cyclus fuel cycle simulator [6] was creating a simulation environment that is more expressive than more traditional fuel cycle simulators. Cyclus was designed with a plug-in style architecture that allows different models for facility types, known as *archetypes*, to be swapped in and out easily. While the reactor archetypes used in this analysis do not implement internal physics calculations, plug-in archetypes such as Bright-lite [7] have been developed for Cyclus that provide varying degrees modeling fidelity.

As a Cyclus simulation walks through each time step, facilities engage in a market-like dynamic resource exchange (DRE). They broadcast requests for material and potential supplying facilities respond with bids. The requesting facilities then express preferences between all their bids, and the DRE then generates a network flow problem that is optimized to resolve which transactions actually occur. The plug-in architecture combined with the DRE creates a powerful and flexible environment for the natural comparison of different modeling paradigms. Cyclus' module plug-in based architecture in concert with its DRE has the ability to model fleet-based facilities. Additionally, Cyclus can model facilities individually with discrete flows enabling the investigation of real-world effects such as competition, reactor cycle staggering, and individual reactor outages.

Cyclus was designed to operate natively in Unix/Linux type environments with flexible and scriptable usage and input file format. Cyclus stores its comprehensive output data in a powerful open-source database format. In addition to information about material flows and facility deployments, Cyclus' single-file database contains the simulation input file, version information about the archetypes used in a simulation, the version of Cyclus used to run the simulation, and the installed versions of Cyclus dependencies. This and other features of Cyclus lend themselves well to automation, tracking research provenance, reproducibility, and large-scale computing.

# II METHODOLOGY

Two particular design choices are selected for investigation: time step duration and facility discretization. A single fuel cycle transition scenario is run in a set of simulations using combinations of the time step duration and facility discretization — four four different cases:

1. *Case MI*: Monthly time steps with Individual reactor modeling
2. *Case MF*: Monthly time steps with Fleet reactor modeling
3. *Case QI*: Quarterly (3-month) time steps with Individual reactor modeling
4. *Case QF*: Quarterly (3-month) time steps with Fleet reactor modeling

These four cases span a spectrum of fidelity with one end being more realistic (i.e., smaller time steps, individual facilities) and the other being more appropriate for scoping studies (i.e., larger time steps, fleet facilities). Conducting this analysis involved developing some additional capability in/around Cyclus. Three primary pieces to the methodology are:

- A fleet-based reactor model.
- A simulation scenario with both initial conditions and transition details.
- Theory and metrics for comparing differences between the four cases.

Because the standard package of archetypes included with Cyclus does not include a fleet-based reactor model, one was necessarily developed. A simulation scenario was created that is roughly based on prior and ongoing work in the Department of Energy (DOE) fuel cycle options campaign [8]. A few phenomena related to the different modeling choices in each of the four cases are identified and used as a basis for comparison. These are described in more detail in the following sections.

### II.A Fleet Reactor Implementation

A fleet reactor archetype was created for Cyclus. Although the modeling of individual, discrete facilities was a motivation for Cyclus, its flexibility makes it relatively straightforward to create plug-in archetypes to model anywhere along the spectrum from individual, discrete facilities to continuous flow fleet facilities. In order to contrast well with the individual, single-facility granularity modeling that is part of the standard Cyclus archetypes, a fleet archetype was created for this work with aggregate facility behavior and continuous flow. Although it would certainly be interesting to compare many intermediate levels of facility discretization (e.g., 1-reactor groups, 10-reactor groups, single-fleet, etc.), only the two extremes of continuous, single fleet and individual reactor modeling are investigated in order to maintain an appropriate scope.

The fleet reactor, although a single entity modeling many reactors, is actually designed to look like many individual reactors to the Cyclus simulator kernel. This "trick" allows other plug-ins for managing facility deployments to transparently work with the fleet reactor without having to know anything about the fleet paradigm. The fleet reactor has several important characteristics described below:

- Reactor capacity is deployed and retired in increments of single reactor units, each capable of producing power $P_r$ (e.g., 1 GWe LWRs). $N_r$ reactors are deployed at any time capable of producing up to $P_r \cdot N_r$ power.
- In the event of fuel shortages, the number of reactors that can operate is modeled as

$$N_o = N_r \cdot \frac{C_{inv}}{C_{cap}}, \tag{1}$$

where $C_{inv}$ is the total amount of fuel in all reactor cores (kg), and $C_{cap}$ is the total fuel capacity for all reactor cores (kg).

- Material is discharged from the fleet's reactor cores continuously (every time step) according to the following equation:

$$D = \frac{B}{L} \cdot N_o = \frac{B}{L} \cdot N_r \cdot \frac{C_{inv}}{C_{cap}} \tag{2}$$

$D$ is the discharge rate (kg per time step), $B$ is the batch size (kg) for a single reactor's batch, $L$ is the cycle length (including refueling time) in time steps. In the event of fuel shortages, the discharge rate is reduced because some reactors are not operating.

- Refueling occurs continuously with as much fresh fuel as is available up to the total amount to fill all reactor cores in the fleet. No extra fuel is kept on-hand — it is all ordered just-in-time.
- Generated power at any time step, $P_G$, is based on the number of operating reactors:

$$P_G = P_r \cdot N_o = P_r \cdot N_r \cdot \frac{C_{inv}}{C_{cap}} \tag{3}$$

- When a reactor in the fleet is retired, a full core of material is discharged — even if there is some fraction of the fleet with unfueled cores.

One characteristic of this fleet reactor implementation is that it has perfect fuel sharing and perfect cycle staggering among the individual reactors it models. The first discharge of spent fuel begins immediately on the first time step of operation rather than after a complete refueling cycle. Fuel shortages also only cause proportional loss of capacity rather than whole-reactor quantized shutdown. The full implementation of the fleet-based reactor is publicly available for download and use [9].

### II.B Scenario Description

The chosen scenario approximates a transition from the current once-through light water reactor (LWR) fleet to a full sodium fast reactor (SFR) fuel cycle within evaluation group 23 (EG-23) in the DOE's Advanced Nuclear Fuel Cycle Options report [10]. Scenario and transition details were patterned after ongoing work by the DOE in their Fuel Cycle Options campaign.

The scenario starts with 100 LWRs and follows an exponential curve with a 1% annual growth rate for 200 years. The decommissioning of the initial 100 reactors is staggered over years 15 to 55. Fresh fuel for LWRs is provided directly from enrichment with infinite supply and throughput. In year 35, fast reactors become available for deployment and no more thermal reactors are built. LWR spent fuel separations begins in year 15 with 2,000 MTHM/yr capacity and increases to 3,000 MTHM/yr in year 25. Fast reactor spent fuel separations and fuel fabrication have infinite capacity. All spent fuel is stored/cooled for 7 years before it becomes available for separation: 5 years for cooling and an additional 2 years to approximate non-instantaneous separations and fabrication times. Figure 1 shows the material flow relationships between the facilities.

In fuel cycle simulation, it is common to use simplified fresh and discharged fuel compositions that are consistent with results from externally performed loading and burnup calculations. Compositions used in this analysis follow this approach and are fixed (i.e. not dynamically computed in-situ). These compositions are the same as those used in similar transition analyses by Hoffman, et al. [8] and Littell, et al. [11] containing $^{238}U$, $^{235}U$, $^{239}Pu$, $^{241}Am$, and small quantities of a few other isotopes as place-holders representing fission products. Exact composition details can be found in the Cyclus input files used [9].

A pre-release of Cyclus version 1.4 was used for this analysis. A standard library of facility plug-ins, Cycamore, is provided with Cyclus. Like Cyclus, a pre-release of Cycamore version 1.4 was used. A custom storage facility archetype was developed because Cycamore did not yet provide one. Code for the fleet reactor and storage archetypes is publicly available and can be downloaded [9]. The standard Cycamore reactor archetype was used for both LWR and SFR reactor types in cases MI and QI. Correspondingly the custom fleet-based reactor archetype described earlier was used for both reactor types in cases MF and QF. Additional configuration for the reactor types is shown in Table I (cycle length includes the refueling outage).

TABLE I: Reactor Facility Parameters

| | LWR | SFR |
|---|---|---|
| Lifetime (yr) | 80 | 80 |
| Cycle length (months) | 18 | 15 |
| Batch size (kg) | 29,565 | 8,025 |
| Batches per core | 3 | 5 |

The described scenario is used as closely as possible in each of the four cases with appropriate adjustments for case-specific differences (e.g., monthly vs. quarterly time steps). Invariants preserved with respect to reactor behavior between all four cases are shown in Table II. Table III shows the computed/selected configuration for both LWR and SFR reactor types that was used in Cyclus input files for all four cases.

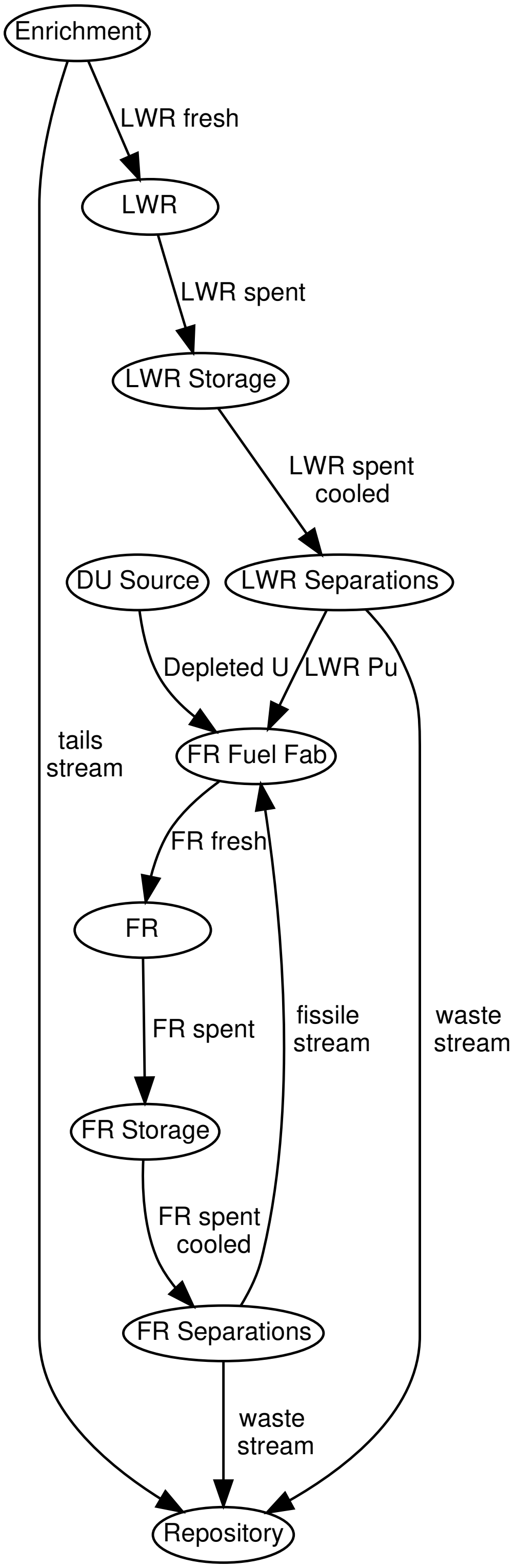

Fig. 1: Material flow paths between facility types. An independent DU source separate from enrichment tails is used for simplicity. The primary fissile isotope in the fissile separations stream is $^{239}Pu$.

TABLE II: Reactor Parameter Invariants

| | LWR | SFR |
|---|---|---|
| Discharge Rate ($\frac{kg{\cdot}HM}{month}$) | 1,642.5 | 535 |
| Burnup ($\frac{MWe{\cdot}month}{kg{\cdot}HM}$) | 0.547945 | 0.672897 |
| Effective Power (MWe) | 900 | 360 |
| Core Size ($kg \cdot HM$) | 88,695 | 40,125 |

TABLE III: Selected Reactor Parameters by Case

| | LWR | | SFR | |
|---|---|---|---|---|
| | Cases MI, QI | Cases MF, QF | Cases MI, QI | Cases MF, QF |
| Cycle length (months) | 15 | 18 | 12 | 15 |
| Refueling outage (months) | 3 | 0 | 3 | 0 |
| Batch size ($kg \cdot HM$) | 29,565 | 29,565 | 8,025 | 8,025 |
| Power capacity (MWe) | 1,080 | 900 | 450 | 360 |

In each of the four cases, the initial 100 LWRs are modified to have a zero-length refueling outage with an explicit 900 MWe net power output capacity. Because the individual reactor model does not have any way to start a reactor mid-cycle, this is a modeling trick that was done to avoid having all initial LWRs refuel at the same time. This adjustment serves to improve the realism associated with the effects being investigated (i.e., fleet-average power generation) even though it is unrealistic in other ways.

The parameters for each case were also selected to keep capacity factors for individual reactors very roughly around 90%, although they are actually closer to 80% because of integer variable constraints on the case MI and QI reactor configuration.

Instead of annual deployments, a longer 21-month deployment period is used because it provides better natural staggering for the reactor refueling cycles. The 630-month lowest common multiple (LCM) for 21-months (deployment period), 15 months (SFR cycle), and 18 months (LWR cycle) is much larger than the 180-month LCM when using a 12-month build period providing a more even power generation profile. The same deployment schedule is used for all four cases; power capacity is built along a curve starting at 100 GWe in year zero following the 1% exponential growth curve. Although many alternate deployment schedules exist that avoid SFR fuel shortages (e.g., a more gradual shift toward SFRs), this particular schedule is deliberately chosen because it causes significant fuel shortages. This is done in order to force the simulation and facility models into more extreme circumstances in order to expose behavioral differences among the design choices investigated.

All input files and other assets used in the analysis for each of the four cases are publicly available for download and use [9].

### II.C Modeling Effects

There are various low-level phenomena that have potential to affect the larger outcome of simulations with respect to facility discretization and time step modeling decisions. Several effects arising from these different modeling choices are present in many fuel cycle simulators to varying degrees. Some of these effects are described in detail below.

One difference between the fleet and individual facility modeling is the *quantized shutdown effect*. If a fleet-based reactor modeling 3-batch cores is short one half batch of fuel, it will produce $\frac{2.5}{3} \cdot P_r$ of power for that time step where $P_r$ is the power capacity of a single reactor of the fleet (e.g., 1000 MWe). If an individually modeled reactor is short one half batch of fuel, then it will produce no power $(0 \cdot P_r)$ that time step. Time step duration can amplify this effect. For both fleet and individual reactor modeling, power goes offline in full time step increments. Consider an example of a 3-batch reactor with a fuel shortage of $\frac{1}{2}$ batch of fuel that would be resolved in 1 month. Depending on the modeling choice, each reactor would produce the following amounts of energy over a 3 month period:

- case *QF*: $\frac{2.5}{3} \cdot 3 \text{ mo} \cdot P_r = 2.5 P_r$
- case *MF*: $[\frac{2.5}{3} \cdot 1 \text{ mo} + 1 \cdot 2\text{mo}] \cdot P_r = 2.83 P_r$
- case *QI*: $0 \cdot 3 \text{ mo} \cdot P_r = 0$
- case *MI*: $(0 \cdot 1 \text{ mo} + 1 \cdot 2 \text{ mo}) \cdot P_r = 2 P_r$

Quantifying this effect directly is difficult. One way of observing it is to compare fleet reactor scenarios that have different time step durations (i.e., cases MF and QF). Because the fleet reactors have perfect fuel sharing (described below) and no noise from refueling outages, differences observed will partly be a result of quantized shutdown.

Individually modeled reactors each have their own refueling cycles. Depending on how the refueling cycles and outages are staggered, power production can vary significantly. For example, if all reactors refuel at the same time, then all reactors are offline at the same time. This is referred to as the *cycle staggering effect* and does not occur in fleet-based reactor models. An example of poor cycle staggering is shown in Figure 2. This effect can be naturally observed as the fluctuations in the difference between deployed power capacity and generated power. The figure shows power jumping above and below the installed net capacity (includes capacity factor) from time step to time step depending on how many reactors are online/offline together.

Poor reactor cycle staggering can also cause spikes of material supply and demand that can lead to suboptimal resource utilization. Consider a fuel fabrication facility that receives requests for new batches for every reactor all at once. The reactors draw out inventory from the fuel fabrication facility together in a large quantity of orders. If the fabrication facility does not have enough material on hand, some number of reactors will need to wait until enough fuel can be fabricated. Avoiding such constraints would require the fabrication facility to maintain suboptimally large on-hand inventories as a contingency. Even with infinite material availability, poor staggering can result in supply constraints caused by finite facility throughput. Even if the fuel fabrication facility above has infinite material supply that it can keep on hand, it might not be able to fabricate fuel for all reactor requests all at once even though it has sufficient capacity if they request fuel uniformly over time.

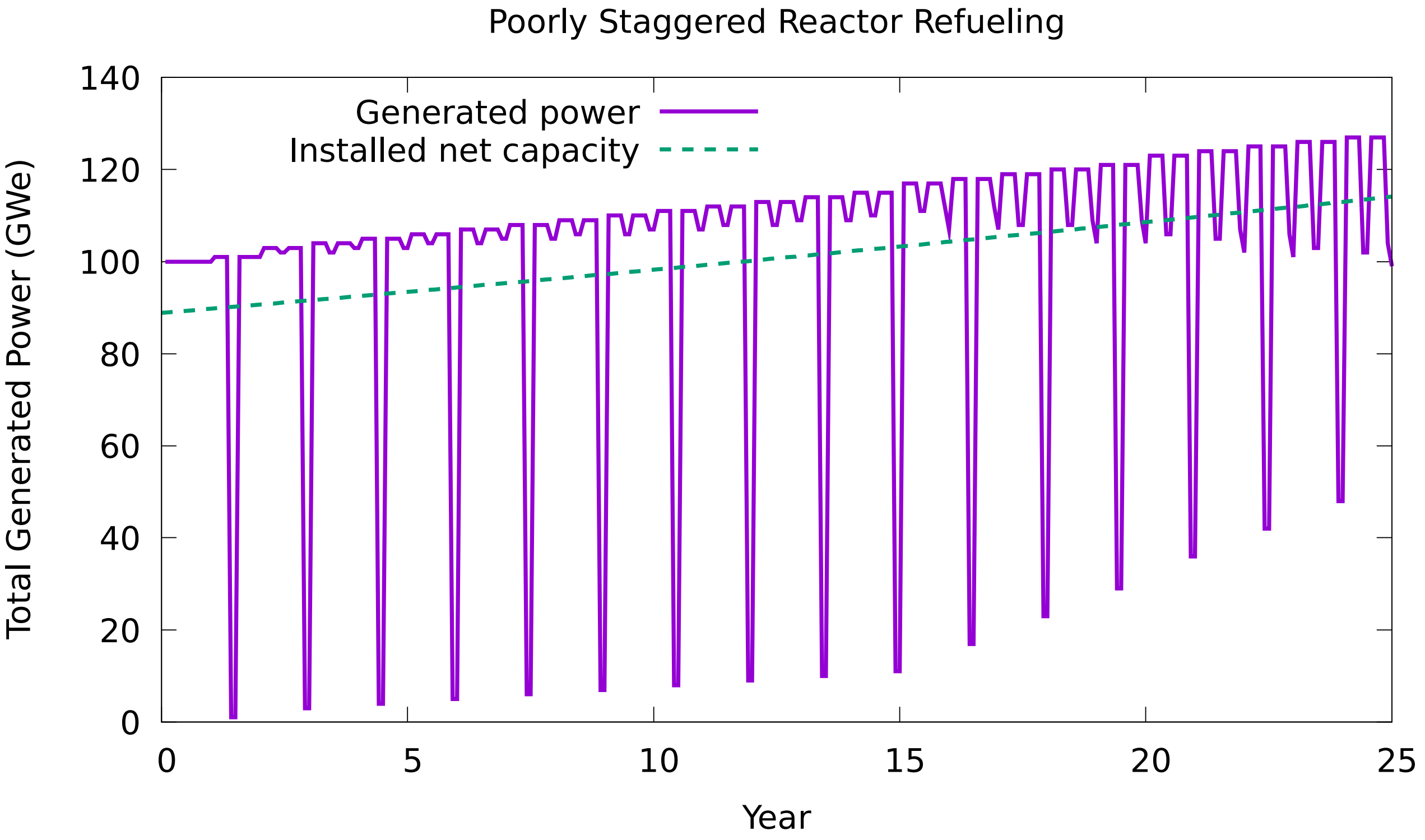

Fig. 2: In this scenario, reactors are deployed annually with a refueling cycle length (including outage) of 18 months. As a result, all reactors deployed every 3rd year refuel together. All initial 100 LWRs start out with their cycles synchronized as well.

Another difference between fleet-based and individual facility modeling involves fuel sharing. Power production by a group of individual reactors for a fixed amount of fuel shortage may vary depending on how available fuel is distributed. This is illustrated in Figure 3. A system being short 3 batches might mean that one reactor has no batches and is not operating or that 3 reactors are each missing a single batch. This is referred to as the *fuel sharing effect*. Fleet reactors, by design, have perfect fuel sharing — all fuel is distributed so that the minimum possible amount of capacity is offline during shortages. The most correct measure of fuel sharing inefficiency is the difference between the actual generated power and the maximum amount of power that could have been generated among all possible fuel distribution alternatives. This is difficult to actually compute. One possible approximation is to count the number of fresh fuel batches distributed to reactors that ended up not having enough fuel to operate (i.e., a full core) that time step. This approximation assumes that every "wasted" batch of fuel, could have been given to a reactor that only needed one batch to operate. Equation 4 describes such an approximation.

$$N_{wasted}(t) = \sum_{r \in R_t} \left[ H[t - (\tau(t,r) + \Delta t_{out})] \cdot [1 - O(t,r)] \sum_{t'=F_{prev}(t,r)}^{t} N_b(t',r) \right] \quad (4)$$

where $R_t$ is the set of all reactors, $H$ is the Heaviside function, $\tau(t,r)$ is the beginning of the most recent refueling outage of reactor $r$ before $t$, $\Delta t_{out}$ is the normal refueling outage duration, $F_{prev}(t,r)$ is the start of the most recent refueling outage for reactor $r$ before or on time $t$, $O(t,r)$ is 1 if reactor $r$ is producing power at time $t$ and zero otherwise, and $N_b(t,r)$ is the number of new fuel batches received by reactor $r$ since time $t$.

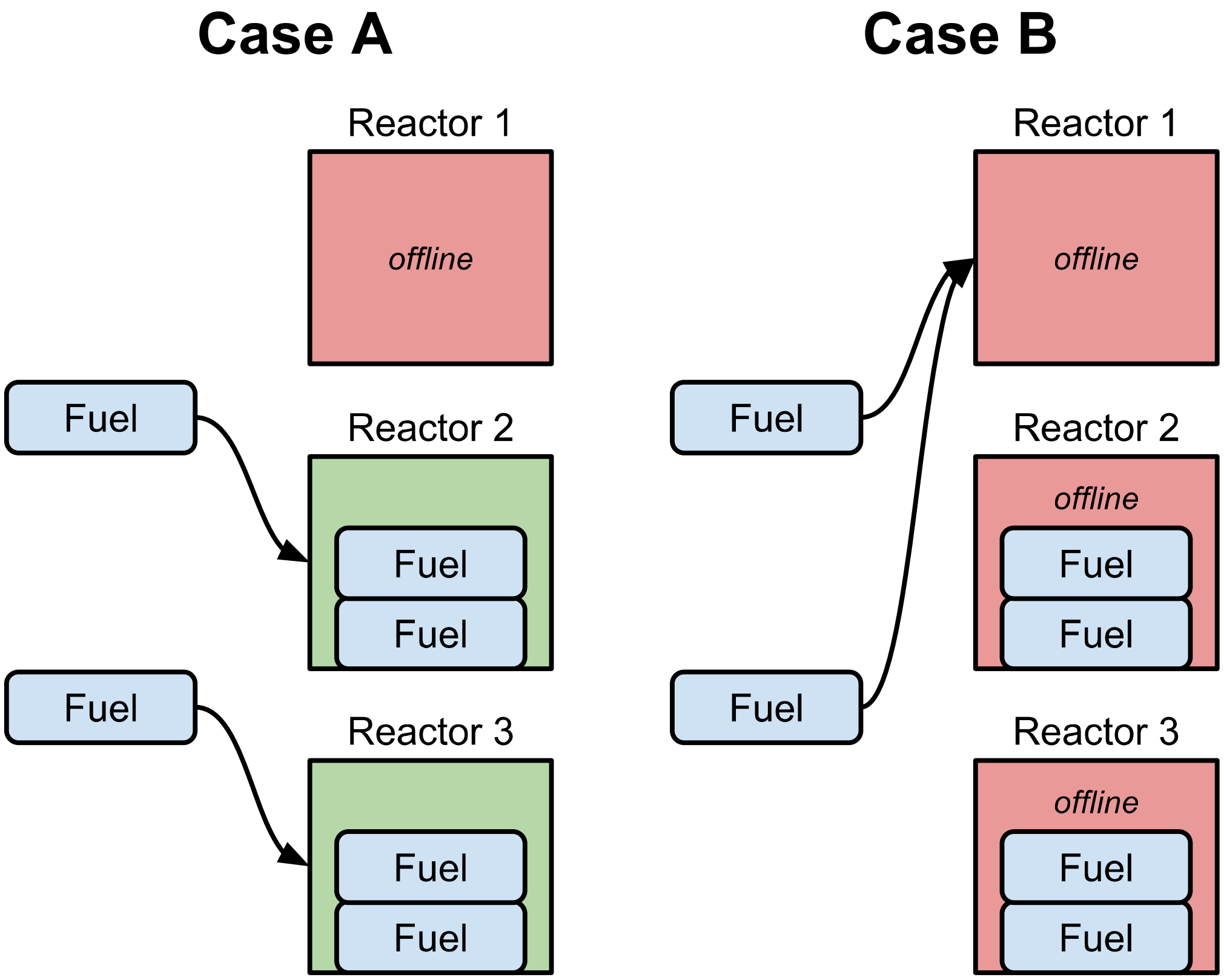

Fig. 3: A simple diagram showing effects of constrained fuel supply distribution choices. In both cases A and B, reactor 1 needs 3 new fuel batches to operate and reactors 2 and 3 each need 1 fresh batch. In case A, the two batches are given to reactors 2 and 3, resulting in reactor 1 remaining offline. In case B, the available fuel batches are given to reactor 1 resulting in all three reactors being offline.

One important phenomenon related to time step duration is the *inventory drawdown effect*. Increasing time step duration reduces the frequency over which refueling can occur, resulting in larger impulse drawdown on inventories. This by itself is not problematic because this is balanced by correspondingly larger inventory top-up quantities. However, another component of this effect is the simultaneity of incoming and outgoing material flows on a given time step. Within a particular time step facilities do not know about incoming material when they make commitments for outgoing material. For longer time steps, larger incoming quanta of material are not available for making offers. In general longer time steps create a need for larger floating inventory buffers in order to avoid material shortages. Another aspect of this effect is that at least one time step is required for a material object to move between two facilities. Larger time steps result in longer minimum bounds on time to traverse paths between facilities. Although direct measures of this effect are difficult, some of its consequences (e.g., higher standing inventories during shortages) can be quantified by comparing scenarios with different time step durations (e.g., cases MF and QF).

There are also performance differences between the two modeling paradigms. A single 200 year simulation with several hundred fleet-based reactors takes about 2 seconds to run on a computer with an Intel Core(TM) i7-4770 3.40GHz CPU where the same simulation using individual reactor facilities takes about 20 seconds. This order of magnitude difference is an important trade-off to keep in mind when doing fuel cycle analysis in general, particularly when considering optimization or sensitivity studies where running thousands or millions of simulations may be desirable. The time step duration has less impact on performance because the amount of work being done in the simulation is similar; in the monthly time step cases most facilities (i.e., all reactors) only perform actions every refueling cycle rather than every time step.

## III RESULTS

### III.A Power Production

Figure 4 shows an overall view of the generated power over time for each of the four cases. Figure 5 normalizes the Figure 4 curves to the expected exponential power curve magnifying some interesting differences between the four cases. For the first 100 years, the four cases behave somewhat similarly, although the individually modeled reactors in cases MI and QI actually go offline and back online for refueling causing more variance. Around year 100, a fuel shortage begins and more significant differences between the four cases become apparent. This fuel shortage lasts until about year 140. The power generated in case QF is slightly lower than in case MF during the fuel shortage years. This is partly a result of the *quantized shutdown effect*. The longer 3-month time step in case QF results in reactor capacity going offline longer than necessary.

During the initial 100 years, cases MI and QI have larger variance in power output than cases MF and QF. This is caused by minor refueling cycle synchronization. The *cycle staggering effect* becomes much more visible during and after the fuel shortage as evidenced by the larger swings in power generation from time step to time step. Initially, the reactor cycles are staggered well. During the shortage, several reactors that previously had staggered cycles are all offline together waiting for fuel. On time steps where reactors retire (not just during the shortage), new deployments are made to replace them in addition to new deployments made to address power demand growth. Because the initial reactors retire in waves somewhat close together, there are corresponding waves of deployment, and these waves are echoed every 80 years (the reactor lifetime). These waves of deployment cause surges in recycled fuel availability when they begin to discharge their fuel that cause many of the reactors offline during the shortage to receive fuel and come online together. The net effect is that the fuel shortage degrades reactor cycle staggering overall. As the simulation continues, however, new reactors continue to be deployed and reactors with synchronized cycles retire resulting in a gradual return to better staggering visible by the end of the 200 years.

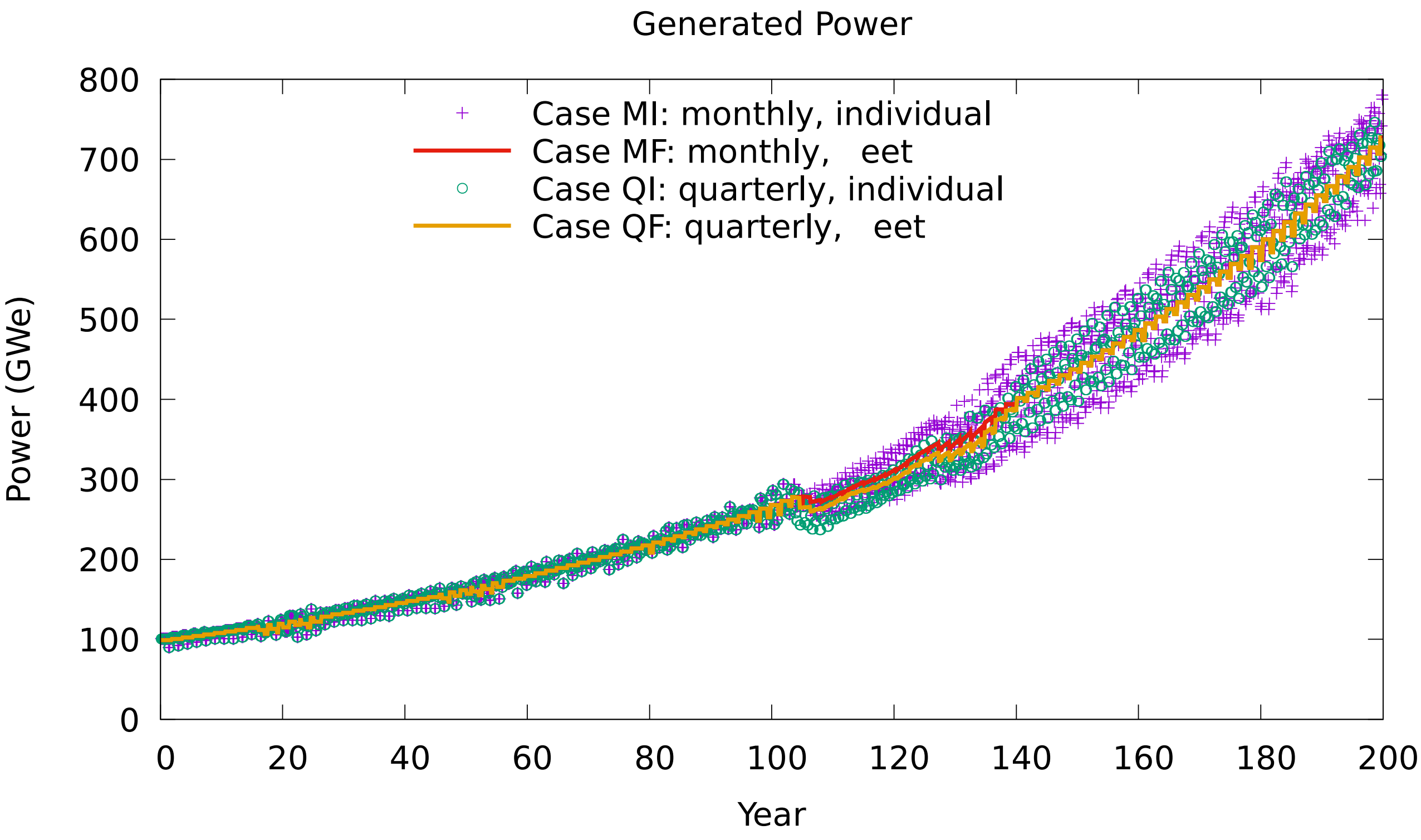

Fig. 4: Generated power for all four cases. A fuel shortage occurs from about year 100 to about year 140. In cases MI and QI with individual reactor modeling, refueling cycles become much more synchronized during and after the fuel shortage resulting in large power fluctuations — an artifact of the *cycle staggering effect*.

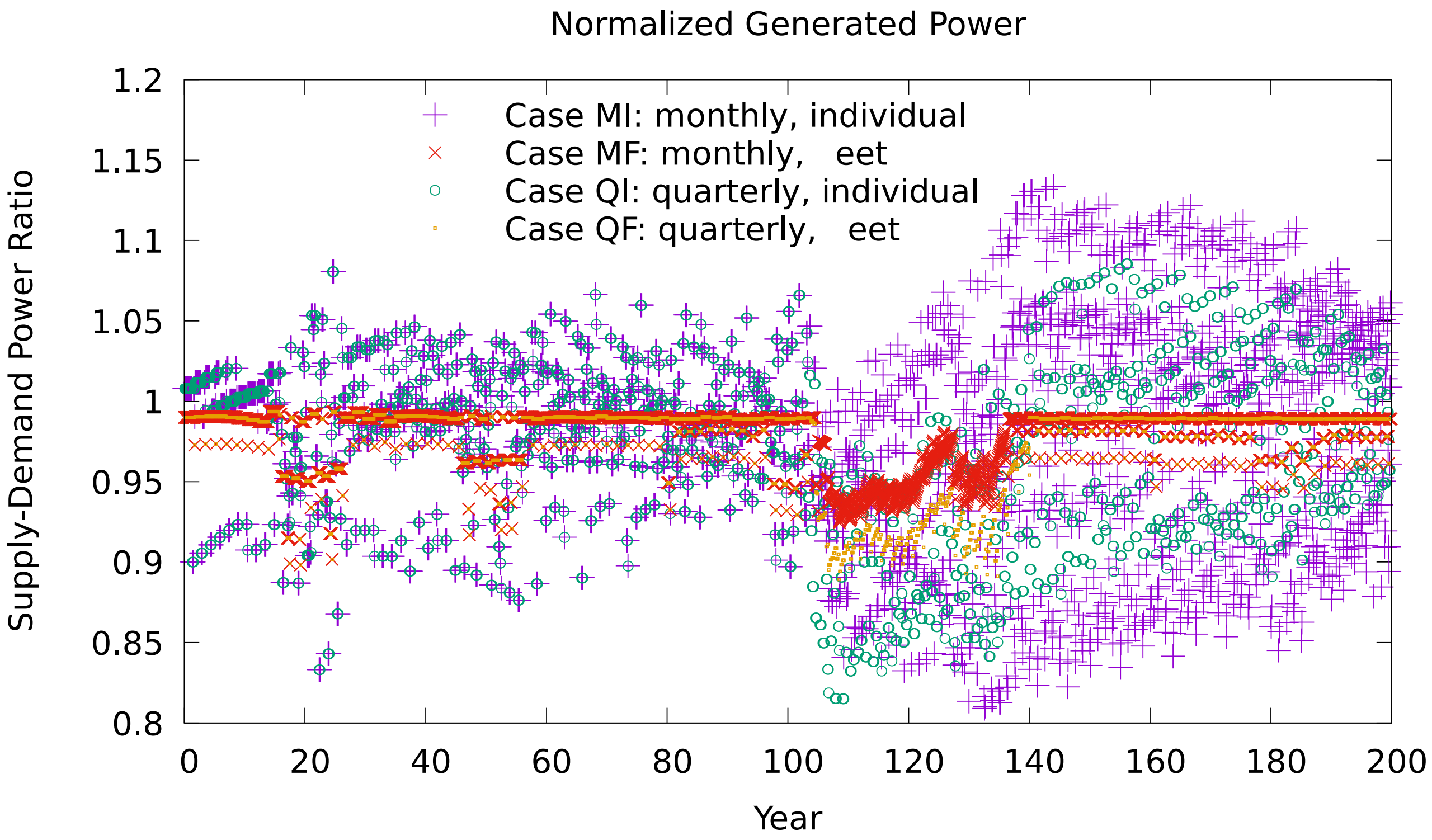

Fig. 5: Generated power is normalized to the expected 1% exponential growth curve for all four cases. A fuel shortage occurs from about year 100 to about year 140. The *cycle staggering effect* can be seen as excessive divergence above and below 1.0 in cases MI and QI where the shortage increases cycle synchronization significantly. The *quantized shutdown effect* causes part of the discrepancy between quarterly and monthly time steps for fleet reactor modeling (cases MF and QF) during the shortage; a longer time step causes more reactor outage than necessary. The consistent jumping up and down for the fleet reactor cases (even pre-shortage) is caused by the misalignment of reactors going offline (multiples of 12 months) and the build period (21 months).

### III.B Fuel Shortage

For the fleet reactors, fuel shortage is exactly the difference between generated power and installed power capacity. Measuring the fuel shortage for individual reactor cases, however, requires careful accounting of the difference between reactors that are offline for a normal refueling outage and reactors that are offline because they have insufficient fuel. Equation 5 shows how this is calculated for cases MI and QI.

$$P_{outage}(t) = \sum_{r \in R_t} P_r \cdot H[t - (\tau(t,r) + \Delta t_{out})] \cdot [1 - O(t,r)] \quad (5)$$

where $R_t$ is the set of all reactors, $P_r$ is the power capacity of reactor $r$, $H$ is the Heaviside function, $\tau(t,r)$ is the beginning of the most recent refueling outage of reactor $r$ before $t$, $\Delta t_{out}$ is the normal refueling outage duration, and $O(t,r)$ is 1 if reactor $r$ is producing power at time $t$ and zero otherwise.

Figure 6 shows the fuel shortage more explicitly for each of the four cases with the case MI and QI results being the cumulative version of Equation 5. The individual reactor modeling cases result in more cumulative outage than corresponding fleet-based cases. The quarterly time step of cases QI and QF also make the outage significantly more severe. Cases MI and QI exhibit approximately twice as much offline power as their corresponding fleet based simulations in cases MF and QF. The monthly time step simulations also have roughly twice as much offline power as their corresponding quarterly cases.

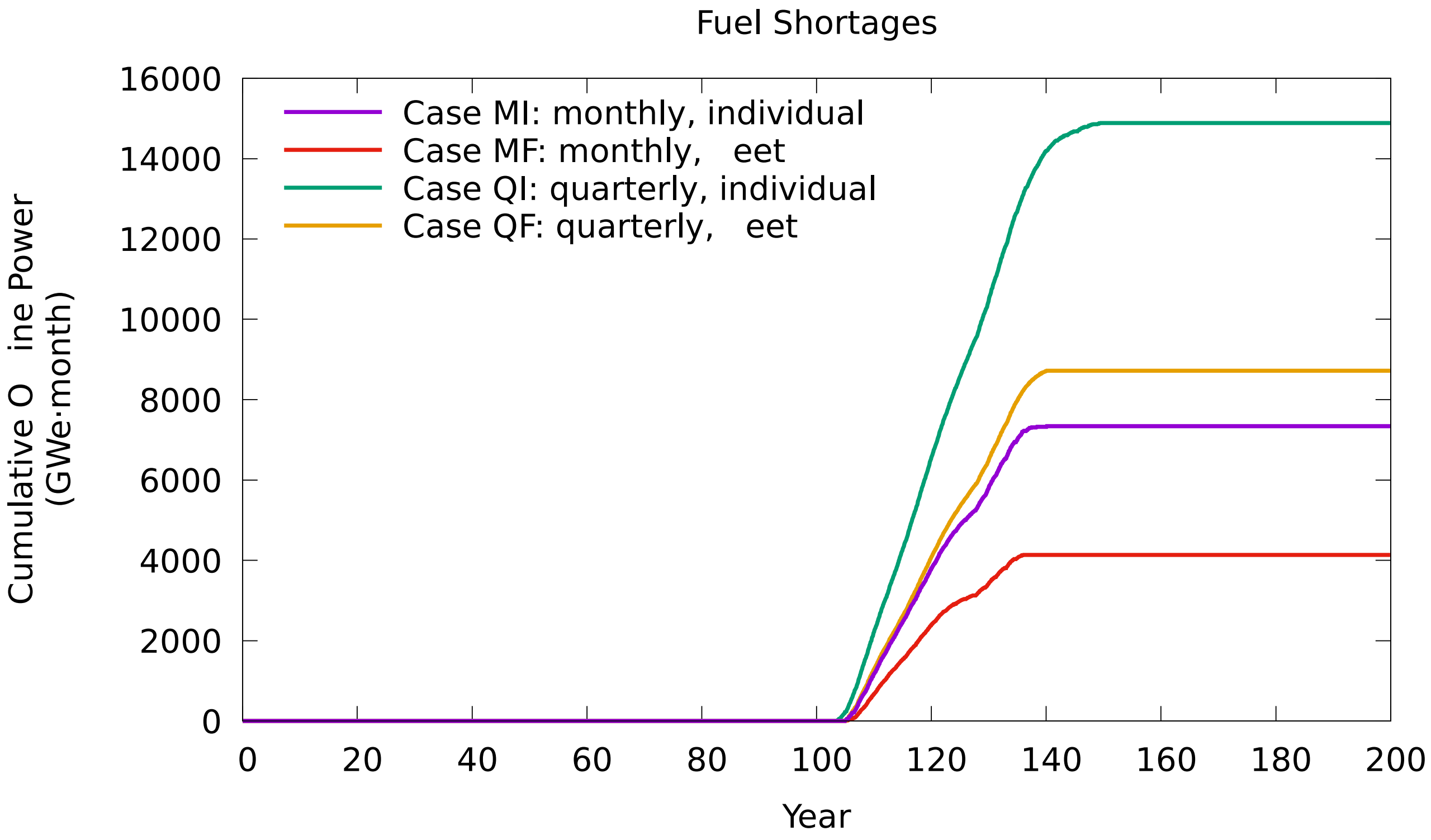

Fig. 6: Cumulative offline power for reactors that had delayed cycle start caused by fuel shortage. The difference between individual and fleet based modeling is primarily a result of the *fuel sharing effect*. The difference between monthly and quarterly time steps is primarily a result of the *inventory drawdown effect*.

Quantifying imperfect fuel sharing is a bit difficult, but Figure 7 provides one way to see the *fuel sharing effect* showing wasted batch-months computed using Equation 4. Fleet reactor modeling, by design, exhibits perfect fuel sharing (zero inefficiency) and so is not included in this figure. Every batch that is given to a reactor that ends up not being able to start its cycle at the scheduled time (caused by fuel shortage) adds one to this cumulative total for each time step the cycle start is delayed.

One useful overestimate of the fuel sharing inefficiency is to assume every one of the "wasted" batches could have been given to a reactor enabling it to operate. Multiplying each of these cumulative batches by the fast reactor power capacity (i.e., 450 MWe) provides a way to compare fuel sharing inefficiency with the overall fuel outage. This results in a cumulative energy deficit caused by poor fuel sharing of roughly 2,000 GWe·months and 6,600 GWe·months for cases MI and QI respectively. These approximations account for a significant portion of the difference between individual and fleet cases in Figure 6 suggesting that most of the discrepancy in cumulative outage between fleet and individual modeling is caused by inefficient fuel sharing in the individual reactor cases.

The poor fuel sharing exhibited in the individual reactor cases would not occur if there were no new reactors being built during the shortage. Previously operating reactors only ever need one batch - receiving only a single batch allows them to operate. Newly constructed reactors need multiple batches in order to begin operation, and if they receive less than a full core they cannot operate. The real world doesn't necessarily optimize for fuel sharing efficiency very cleanly. There are many factors that can affect real-world fuel sharing outcomes. A few include:

1. Multiple fuel quanta comprise a single batch (i.e., multiple assemblies per batch). This functionality is natively supported in the individual reactor model.

2. On-hand fresh fuel inventory maintained at reactors. This functionality is natively supported in the individual reactor model.

3. Long-term fuel contracts between reactors operators and fuel suppliers.

Having multiple assemblies per batch (item 1 above) will further degrade fuel sharing efficiency. Even reactors that are short only one batch could potentially receive some fuel and still be unable to operate. In order to maintain optimal sharing, fresh fuel now needs to be sent in all-or-nothing multi-assembly quanta with higher preference to reactors that need fewer assemblies. Spare fresh fuel inventory (item 2 above) will have the global effect of increasing total shortage severity; on average more batches of fresh fuel will be idling unused. However, it can potentially reduce the frequency of outages for individual reactors in some cases.

If they were to ever occur, real world fuel shortages would likely not result in optimal fuel sharing. However, modeling these the causes of suboptimal fuel sharing is probably best accomplished with a more direct, intentional approach rather than as a modeling artifact as seen here in cases MI and QI. One way to alleviate the poor fuel sharing is to modify the individual reactor model to adjust the preference value on requests for fresh fuel depending on how many assemblies it needs to have a full core. The more assemblies it needs, the lower it will set its request preference. This will have the effect of allowing the DRE in Cyclus to prefer sending fuel to reactors that need less to operate.

### III.C Inventory Drawdown

The large differences seen between monthly and quarterly time step cases in Figures 6 and 7 are primarily a result of the *inventory drawdown effect*. Figure 8 helps to visualize this effect. The black dots in the figure represent impulse flows of available Pu from fuel fabrication and into fast reactors. Blue lines represent the flows into Pu inventory available for making fuel. The in-flows are not shown for the individual reactor cases because they are somewhat messy and obscure other information in the plots. The other colored curves show Pu inventory available for fabricating fast reactor fuel. When the inventory curves meet and force down the out-flows, fuel shortages occur. While this is most clear for cases MF and QF, the individual reactor scenarios (cases MI and QI) also have many points during the shortage where the outflow point lies exactly on top of the inventory point — indicating that all available inventory is being transferred.

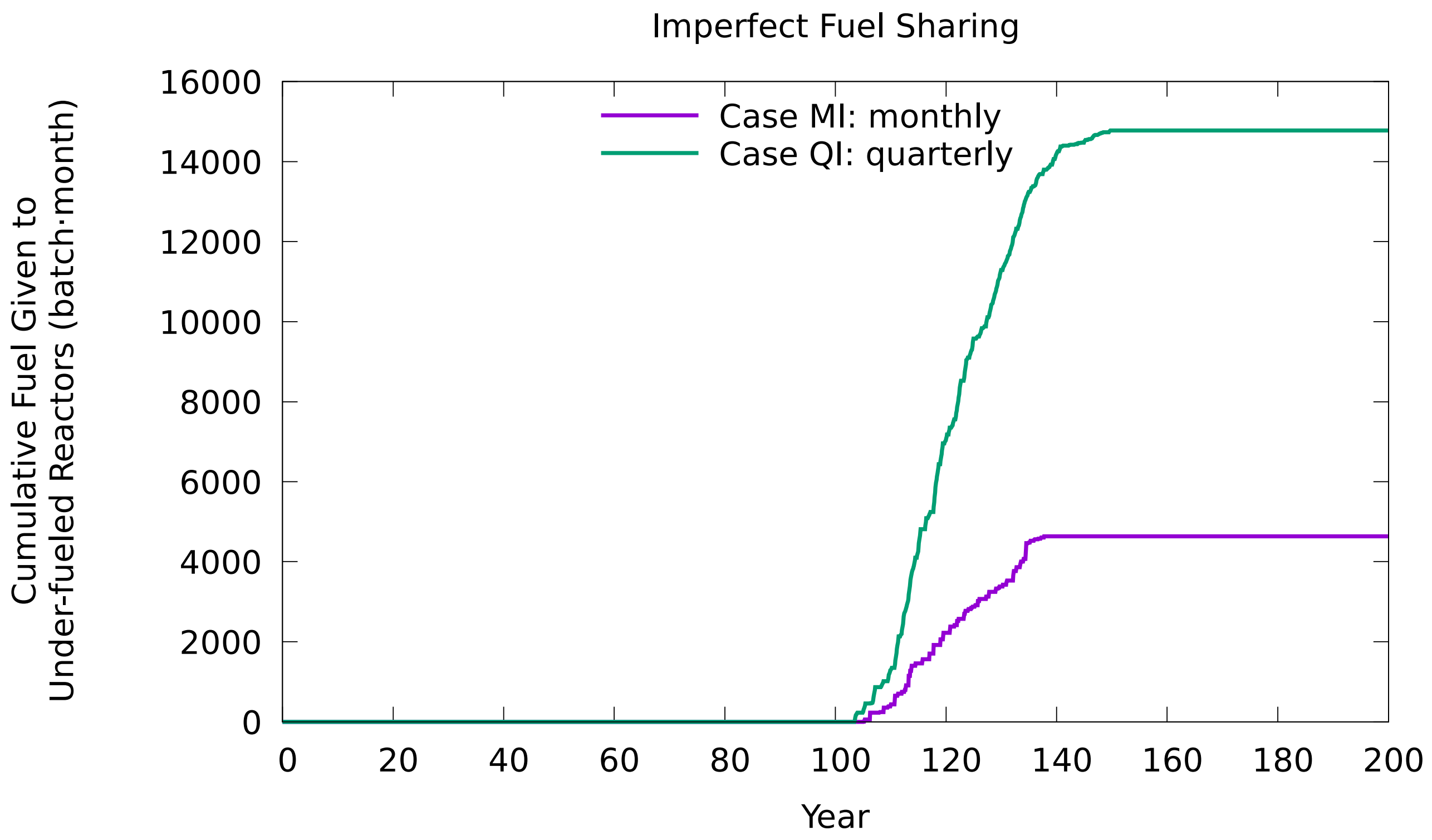

Fig. 7: Fuel sharing inefficiency approximated by the cumulative number of fuel batches given to each reactor on each refueling period multiplied by the number of time steps that reactor had to delay the start of its next cycle (caused by fuel shortage). Fleet reactors implicitly have perfect fuel sharing and so are not included here. The large difference between cases MI and QI is mostly a result of the *inventory drawdown effect*.

Quarterly time steps have less frequent but larger impulse material flows. The larger in-flows, however, do not compensate for the larger withdrawals because facilities do not know about incoming inventory when they resolve their outflow for a particular time step. This information lag effectively requires the floating Pu inventory to be maintained at a higher level during the shortage. This can be clearly seen in Figures 8 and 11 where case QF has a higher standing inventory than case MF during the fuel shortage years. The case QF Pu inventory is unable to drop below approximately 100 tonnes, which indicates out-flows to be roughly equal to the Pu flow into inventory each time step during those years. This can be confirmed by the in-flow and inventory curves touching at those points. Case MF Pu inventory is similarly limited at about the 30 tonne level. As expected, the case QF withdrawals are also three times higher than the case MF withdrawals, exactly matching the time step duration ratio between the two cases.

Throughout most of the shortage, the aggregate outflow of Pu from separated inventory is roughly the same in cases MF and QF. However, because of the longer time steps in case QF, the shortage starts slightly sooner because Pu inventory must be maintained at a higher limit as described above. The small difference in outage start time has feedback effects. Not only are extra reactors offline, but those extra reactors are also discharging less aggregate spent fuel for recycling. This increases the discrepancy between the two cases. Eventually the shortage in case MF ends slightly before it does in case QF (shown most clearly in Figure 12). This feedback effect occurs because the reactors have a conversion ratio greater than 1.0.

The horizontal striations visible in the monthly flows for case MI in Figure 8 are a characteristic of the synchronization of reactor refueling caused by the shortage. A close-up view of these can be seen in Figure 9. They begin at the end of the fuel shortage window when many reactors come online together in groups and are resonances that the inventory level jumps between. This artifact is associated with poor cycle staggering like that shown earlier in Figure 2. These resonances gradually disappear as the simulation progresses farther in time beyond the shortage and refueling cycles become more staggered.

The in-flows shown for cases MF and QF show a periodic pulsing that begins during the shortage in about year 120 and is shown in greater detail in Figure 10. This pulsing actually has nothing to do with the shortage and is caused by the retirement of fast reactors. The first fast reactors are deployed in year 35 and have an 80 year lifetime. When they retire, they discharge an entire core's worth of fuel rather than a single batch. After a 7-year cooling period, the first discharged full cores begin to make it through the recycle loop in year 122. Another interesting feature visible in all four cases in Figure 8 is the minority fraction of out-flow points that are higher than the bulk majority. These higher out-flows occur once every 21 months when new reactors are built because new reactors require a full core of fresh fuel rather than the single batch needed when just refueling.

Figure 11 shows the same inventory curves from Figure 8 superimposed. Perhaps counter-intuitively, lower separated plutonium inventory during the shortage generally indicates fewer unfueled reactors. Lower inventory levels mean available Pu is being utilized more efficiently rather than sitting idle. Cases QI and QF with their longer time steps show higher inventories indicating a more severe shortage. Cases QI and QF take longer to recover from the shortage and fall a bit behind with respect to building up Pu stocks because of the delayed contributions of more unfueled reactors to the separated Pu pool. This can be seen most clearly in Figure 12. Cases MI and QF recover Pu inventory post-shortage at about the same time and rate — this is consistent with their approximately equivalent cumulative offline power capacity curves in Figure 6.

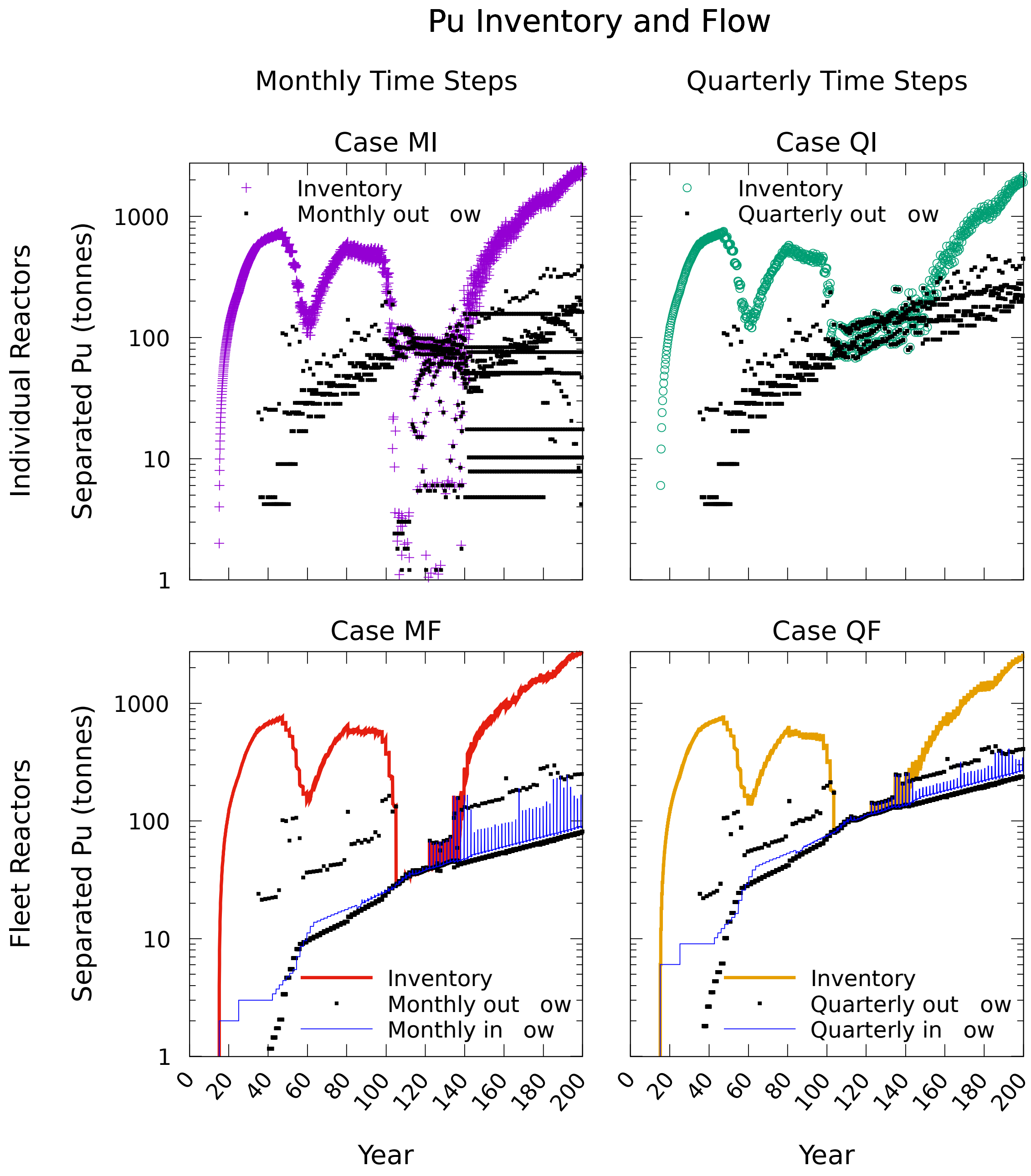

Fig. 8: Separated Pu inventory and flows for all four cases. In-flows for cases MI and QI are omitted because they are noisy and obscure other useful information. When out-flows meet inventory, fuel shortages are taking place. A larger time step results in larger per time step material flows. These larger flows mean more material is not available for supplying to requesters. This information lag causes more supply constraints than occur with smaller time steps - illustrating the *inventory drawdown effect*. Another consequence of this effect is the larger standing Pu inventory during the shortage with larger time steps that can be seen in case QF than in case MF.

The reactors in all the scenarios request fresh fuel only when they need it. While this is an unrealistic behavior for real-world facilities, it is closer to a globally optimum fuel management strategy. Keeping on-hand fresh fuel batches would also have the same effect of increasing idling Pu quantities and would make the fuel shortage worse overall.

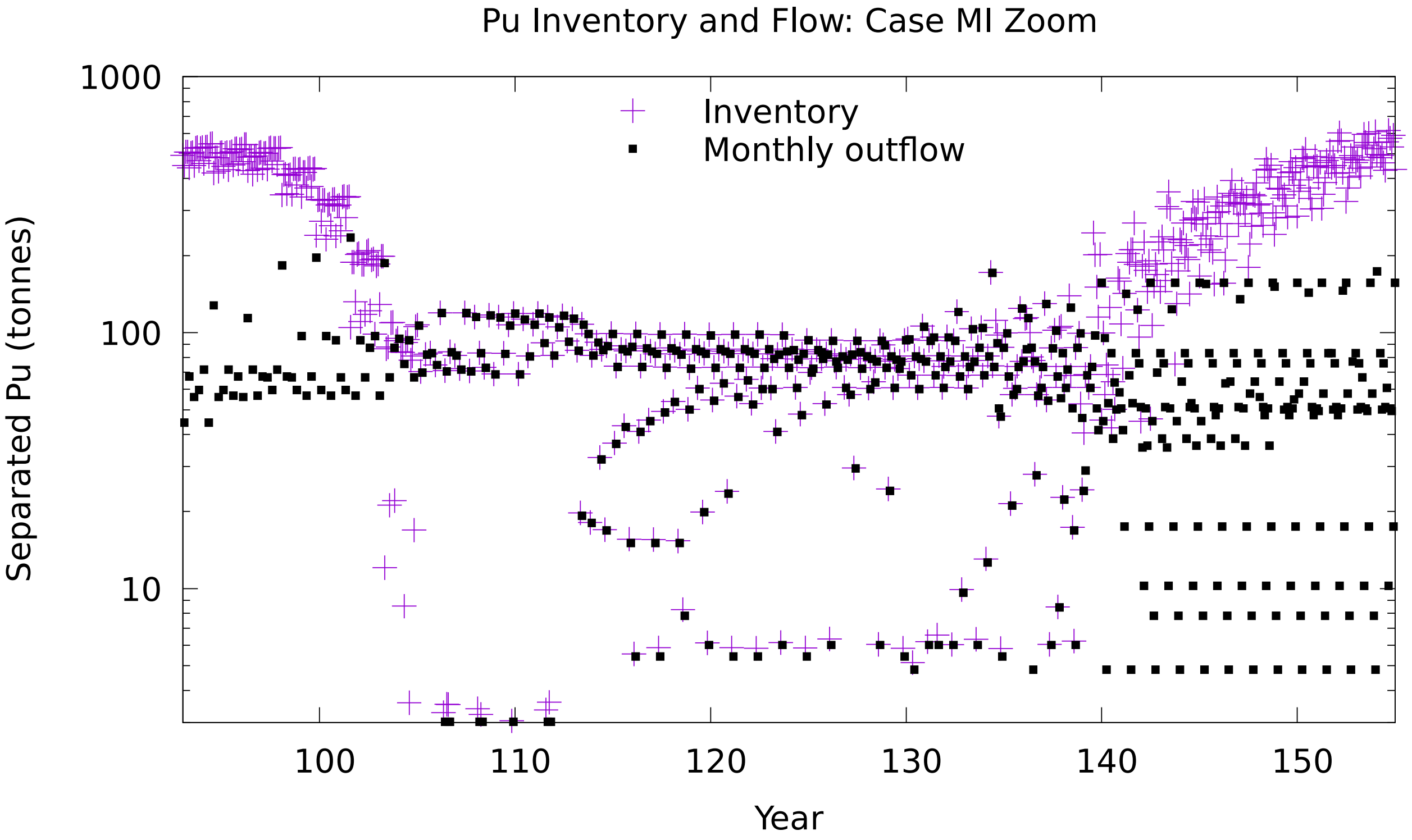

Fig. 9: Zoom view of separated Pu inventory and flows from Figure 8 for case MI. At locations where the outflow dots lie on/near the inventory level, fuel shortages are taking place. This begins around year 105 and continues until about year 140 where Pu withdrawals become lower than standing inventory. The horizontal bands in out-flows starting around year 140 are caused by many previously offline reactors coming online together in groups. These fuel inventory level resonances gradually disappear as the simulation progresses.

Also notable is that different modeling choices (e.g., time step duration, facility discretization, etc.) may have varying levels of significance depending on the scenario/context they are operating in. As shown in Figure 11 before the fuel shortage begins, differences between each of the four cases are relatively minor. However, after the shortage begins near year 100, separated Pu inventories vary much more significantly between the different cases.

Raw data, custom code, instructions for reproducing results, and other artifacts are available for download and use at http://dx.doi.org/10.6084/m9.figshare.1546775[9].

## IV CONCLUSION

Different modeling choices can have a significant impact on the outcome of a simulation. With discrete reactor modeling many factors must be considered in order to ensure the integrity of conclusions.

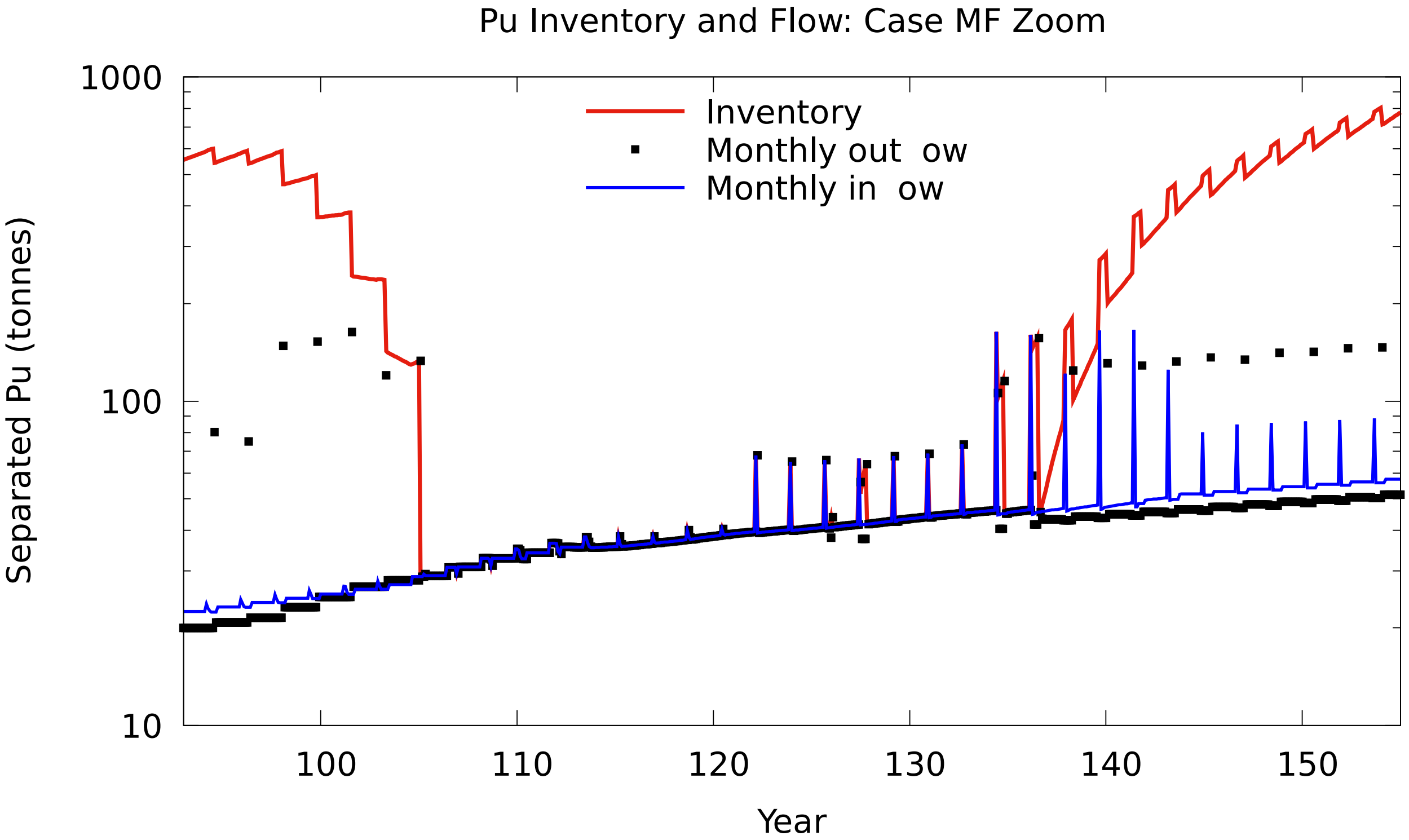

Fig. 10: Zoom view of separated Pu inventory and flows from Figure 8 for case MF. At locations where the outflow dots meet the inventory level, fuel shortages are taking place. This begins around year 105 and continues until about year 140 where Pu withdrawals become lower than standing inventory. The periodic pulsing that begins just after year 120 is caused by the first fast reactor retirements; the first fast reactors deployed starting in year 35 retire annually after their 80 year lifetime and discharge an entire core's worth of fuel rather than the usual single batch for refueling.

Individual reactor outage modeling can be very useful for certain analyses, but it can reduce result quality if things such as cycle staggering are not handled appropriately. The fleet and discrete models used here are just two points in a multi-dimensional spectrum of modeling choices. For example, if fuel sharing is of interest but cycle staggering is not, individual reactors could be used with the power capacity lowered and the outage reduced to zero duration, including the capacity factor in the reactor's power capacity.

Modeling reactors individually can provide valuable insight not possible with the fleet-based reactor modeling. For example, bringing reactors back online following shortage-induced outages should not necessarily follow the natural pacing of fuel availability, otherwise refueling cycles become unsatisfactorily synchronized. If individual reactor modeling fidelity is not required, then performance benefits in both simulation run time and data analysis may suggest a fleet-based modeling approach.

Increasing time step duration increases facilities' standing inventory requirement during supply constrained periods. Here we observe a factor of three difference between standing inventories of separated Pu for a large portion of the simulation — something certain analyses are very sensitive to. Facilities can provide no more inventory than they can hold on a single time step. So increasing time step duration for a facility with unconstrained supply generally means higher standing inventories are required in order to achieve equal throughput. Facilities with large inventory buffer that are unconstrained are not affected significantly by the time step duration when operating in isolation. However, even when facilities do not have individual inventory or throughput limits, the time step can still have a significant effect. The *inventory drawdown effect* can impact the aggregate throughput of supply-constrained recycle loops by affecting minimum bounds on idling material quantities.

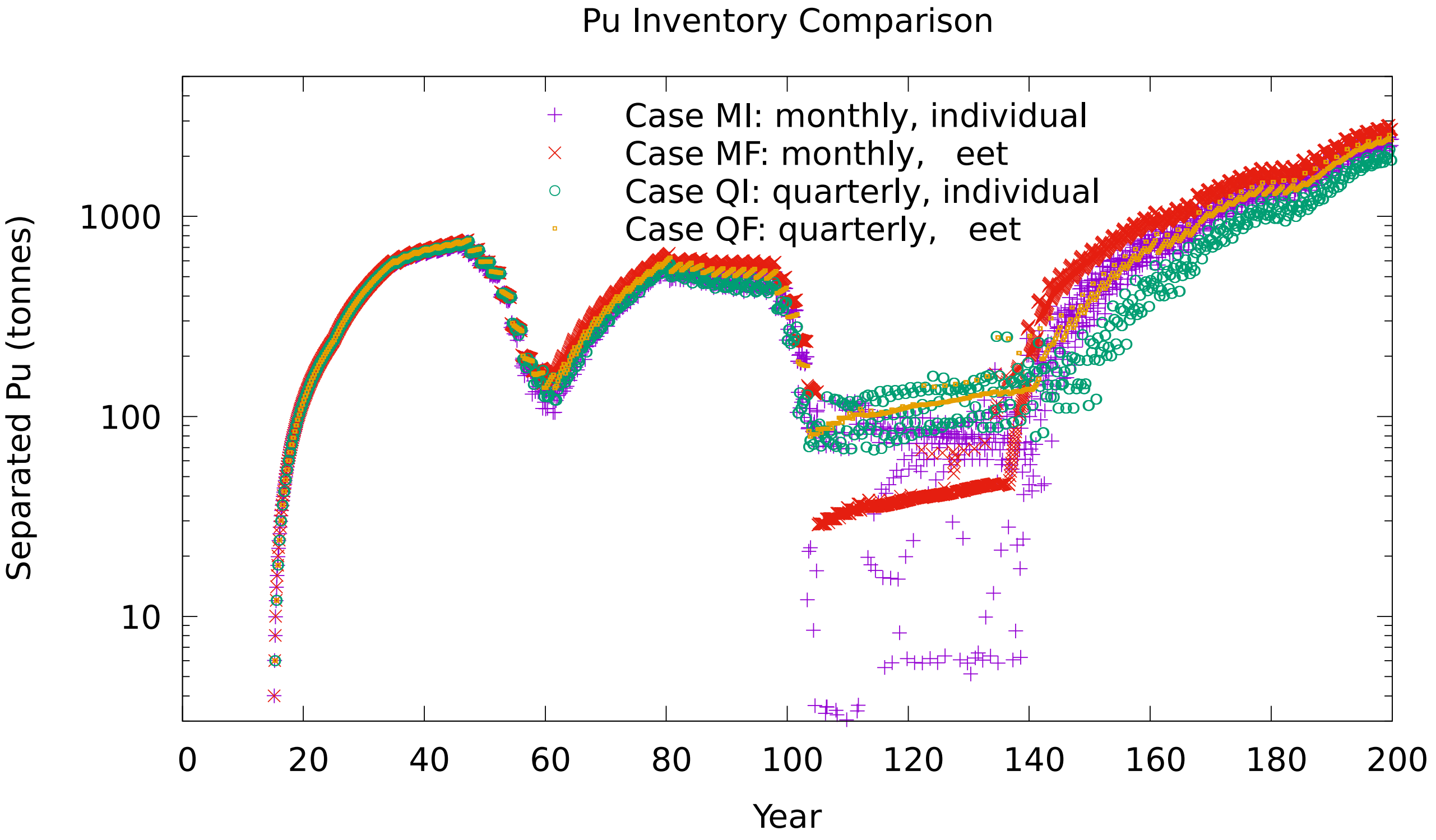

Fig. 11: Separated Pu inventories for all four cases. Differences in modeling choices are not particularly significant until about year 100 when fuel shortages begin. These differences gradually decay away starting in year 140 after fuel shortages have ebbed.

The impact of modeling assumptions depends heavily on the metrics of interest. For example, someone interested in fuel-shortage driven reactor outages would observe a factor of four span among the four cases shown in this comparison exercise. However, someone more interested in the overall transition mechanics would not see as significant a discrepancy. In an unconstrained supply setting, varying the time step duration or discretization of facility modeling has a smaller impact on the overall simulation results. Differences become more significant in material constrained regimes. Robustness in constrained regimes can be especially useful in the context of sensitivity and optimization analysis.

The value of realism in fuel cycle modeling is greatly enhanced by an associated understanding of how outcomes are affected by added fidelity. Cyclus' flexibility for accommodating different modeling choices uniquely enables many interesting comparisons. Those in the field of fuel cycle analysis in general should be cognizant of how modeling choices are affecting results and conclusions through exercises like the one here. Even if the outcome is that a particular modeling decision has little impact on results, efforts to measure and document these impacts can provide confidence in the appropriateness of modeling choices, reducing the need to rely on intuition for cycle analysis work.

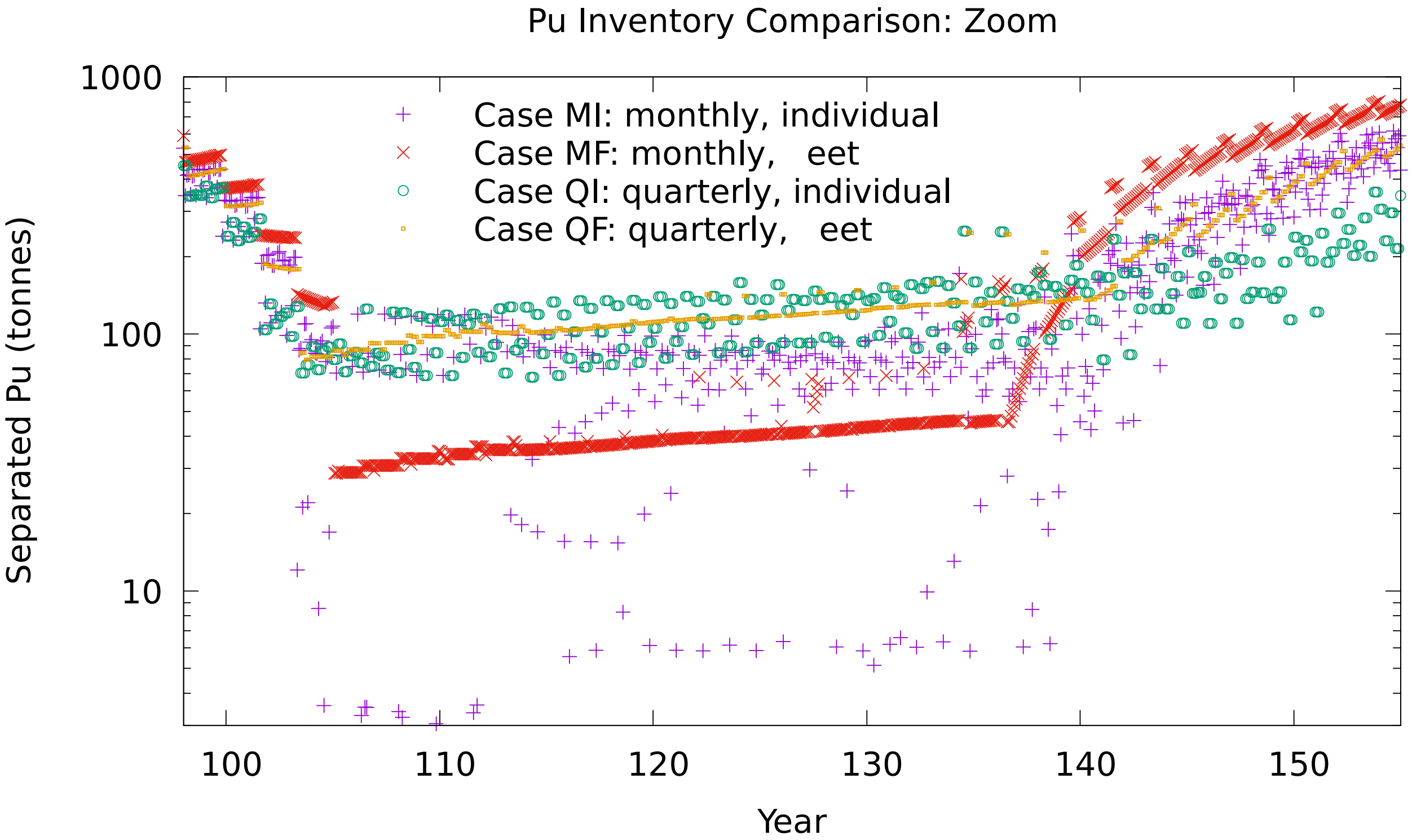

Fig. 12: Separated Pu inventories for all four cases zoomed in on the fuel shortage years. Higher aggregate Pu inventory during the shortage is correlated with both a longer shortage and a slower post-shortage recovery of Pu inventory.

## V ACKNOWLEDGEMENTS

This research is being performed in part using funding received from the DOE Office of Nuclear Energy's Nuclear Energy University Programs. The authors thank the NEUP for its generous support.

NEUP Nuclear Energy University Programs
U.S. Department of Energy

## REFERENCES

1. K. A. MCCARTHY et al., “Benchmark Study on Nuclear Fuel Cycle Transition Scenarios-Analysis Codes,” , Organisation for Economic Co-operation and Development (OECD) (2012).
2. J. W. FORRESTER, *Industrial dynamics*, volume 2, MIT press Cambridge, MA (1961).
3. J. JACOBSON et al., “VERIFIABLE FUEL CYCLE SIMULATION MODEL (VISION): A TOOL FOR ANALYZING NUCLEAR FUEL CYCLE FUTURES,” *Nuclear Technology*,**172**, 157 (2010).
4. V. D. DURPEL, A. YACOUT, D. WADE, T. TAIWO, and U. LAUFERTS, “DANESS V4.2: Overview of Capabilities and Developments,” *Proc. Global 2009*, Paris, France, 2009.
5. R. W. H. GREGG and C. GROVE, “Analysis of the UK Nuclear Fission Roadmap Using the ORION Fuel Cycle Modelling Code,” *Proc. IChemE Nuclear Fuel Cycle Conference*, Manchester, United Kingdom, 2012.
6. K. D. HUFF et al., “Fundamental Concepts in the Cyclus Nuclear Fuel Cycle Simulation Framework,” *Advances in Engineering Software* (2016), Available at http://arxiv.org/abs/1509.03604.
7. R. FLANAGAN, “Bright-lite: A physics-driven reactor facility archetype for the Cyclus framework.,” 2014, https://github.com/bright-dev/bright-lite.
8. E. HOFFMAN et al., “Expanded Analysis of Transition to an Alternative Fuel Cycle,” *Proc. ICAPP 2016*, San Francisco, CA, 2016.
9. R. CARLSEN, “Data for: Challenging Fuel Cycle Modeling Assumptions,” (2015), http://dx.doi.org/10.6084/m9.figshare.1546775.
10. T. WIGELAND, R. TAIWO et al., “Nuclear Fuel Cycle Evaluation and Screening - Final Report,” INL/EXT-14-31465, Idaho National Laboratory (2014).
11. J. LITTELL, S. SKUTNIK, A. WORRALL, and E. SUNNY, “Assessment and Benchmarking with ORION and Cyclus for U.S. Fuel Cycle Options,” *Proc. Global 2015*, Paris, France, 2015.